\documentclass[12pt,preprint]{aastex}

\newcommand{\etal} {et al.\ }

\newcommand{\ie} {{\it i.e.\ }}

\slugcomment{ }

\shorttitle{Dust absorption and the cosmic UV flux density}
\shortauthors{Massarotti, Iovino and Buzzoni}

\begin{document}

\title{Dust absorption and the cosmic ultraviolet flux density}

\author{M. Massarotti\altaffilmark{1}}
\affil{Osservatorio Astronomico di Capodimonte, Via Moiariello 16, 80131 Napoli, Italy}
\email{massarot@brera.mi.astro.it}
\author{A. Iovino}
\affil{Osservatorio Astronomico di Brera, Via Brera 28, 20121 Milano, Italy}
\email{iovino@brera.mi.astro.it}
\and
\author{A. Buzzoni\altaffilmark{1}}
\affil{Telescopio Nazionale Galileo, A.P. 565, 38700 Santa Cruz de La Palma (Tf), Spain}
\email{buzzoni@tng.iac.es}

\altaffiltext{1}{Osservatorio Astronomico di Brera, Via Brera 28, 20121 Milano, Italy}

\begin{abstract}
 We study the evolution of the galaxy UV luminosity density as a
 function of redshift in the Hubble Deep Field North (HDF--N).  We
 estimate the amount of energy absorbed by dust and hidden from
 optical observations by analyzing the HDF--N photometric data with the
 spectral energy distribution fitting method.  According to our
 results, at redshifts $1 \leq z \leq 4.5$, the global energy observed
 in the UV rest--frame at $\lambda=1500$ \AA\ corresponds to only 7-11
 \% of the stellar energy output, the rest of it being absorbed by
 dust and re--emitted in the far--IR. Our estimates of the comoving
 star formation rate density in the universe from the
 extinction--corrected UV emission are consistent with the recent
 results obtained with Submillimeter Common-User Bolometer Array
 (SCUBA) at faint sub--millimeter flux levels.
\end{abstract}

 \keywords{galaxies: distances and redshifts --- galaxies: evolution
 --- ultraviolet: ISM --- dust, extinction --- methods: data analysis
 --- techniques: photometric}

\section{Introduction}

The study of galaxy star formation rate density (SFRD) as a function
of redshift is of crucial importance to exploring cosmic evolution of
galaxies at different epochs and assessing the physical processes
involved in galaxies formation.

Ultraviolet luminosity in star--forming galaxies relates directly to
the number of short--lived high--mass stars, and this gives, in
principle, a measure of the actual SFR.

This simple approach has been widely used for high--redshift objects
such as Lyman--break galaxies, for which rest--frame UV emission can
be probed at optical wavelengths (e.g., Connolly \etal 1997, hereafter
C97; Madau, Pozzetti, \& Dickinson 1998, MPD98; Meurer, Heckman, \&
Calzetti 1999, MHC99; Steidel \etal 1999).

Even moderate amounts of dust may however significantly suppress the
UV flux and, hence, the inferred SFR. A confident assessment of this
effect is required in order to correctly interpret high-redshift data
and recover the cosmic UV luminosity density as a function of redshift.

Meurer \etal (1997) and Sawicki \& Yee (1998), in their analysis of
spectroscopically confirmed $z >2$ galaxies, found that dust can
suppress UV luminosity by a factor $15\div20$ (see also Papovich,
Dickinson, \& Ferguson 2001). Pettini \etal (1998) obtained a factor
about $2.5\div6.3$ from a sample of five Lyman-break galaxies. This is
close to the value of about 5 recently proposed by MHC99 and Steidel
\etal (1999) using their $U$ drop-out samples.

A firm appraisal of the role of dust would also improve estimates of
the contribution of optically--selected starburst galaxies to the
sub--millimeter background, as detected by Submillimeter Common-User
Bolometer Array (SCUBA; see e.g., Adelberger \& Steidel 2000, and
Barger, Cowie, \& Richards 2000, hereafter BCR00).

In this letter we will estimate the comoving luminosity density in the
Hubble Deep Field North (HDF--N, Williams et al. 1996) taking into
account dust obscuration.  As explained in Sec.\ 2, our procedure
relies on the Spectral Energy Distribution (SED) fitting method to
estimate photometric redshift and color excess for each HDF--N galaxy
(see Massarotti, Iovino \& Buzzoni 2001a, hereafter Paper I).  As we
will see, any simplified treatment based on mean corrections for
galaxy $E(B-V)$ could lead to a significant underestimate of the UV
flux density (Sec.\ 3 and 4) and the inferred cosmic SFRD, compared
with the corresponding estimates from the far--IR/submillimeter
spectral window (Sec.\ 5).

\section{The SED fitting method}

For our analysis we use the multicolor photometric catalog of the HDF--N
as obtained by Fern\'andez-Soto, Lanzetta, \& Yahil (1999).  It
consists of 1067 objects observed with the Hubble Space Telescope
(HST) WFPC2 at four photometric bands, $U_{300}$, $B_{450}$, $V_{606}$,
$I_{814}$, and with the IRIM camera at KPNO by Dickinson (1998) in the
$J,H,K$ infrared range.

Photometric redshifts and other properties of galaxies in the sample
have been reported in Paper I and involved fitting the SED, as derived
from the seven-band photometry, with different sets of reference
galaxy models.  The starburst models by Leitherer \etal (1999, L99)
with metallicity $Z = Z_\odot,~Z_\odot/20$ have been complemented by
theoretical templates for irregular, spiral and elliptical galaxies at
old and intermediate ages according to the population synthesis codes
of Buzzoni (1998, 1989, hereafter BUZ) and Bruzual \& Charlot (1993,
hereafter BC; we use the 1998 version code).  We include in the
library also simple stellar population models (by L99) to minimize
confusion between dust absorption and aging effects.  Further details
about the best-fit procedure, as well as the preliminary results of
its application to the photometric redshift technique can be found in
Paper I.

The Calzetti (1999) dust attenuation law has been adopted in our
calculations, including grain absorption and scattering. The $E(B-V)$
parameter was assumed to vary in the range $0.0\div0.4$ mag (we did
not consider higher values of $E(B-V)$, in order to stay within the
range of Calzetti's calibration).  Treatment of the intergalactic
medium absorption as a function of redshift follows from Madau (1995)
and Scott, Bechtold, \& Dobrzycki (2000; see Massarotti \etal 2001b).

The match with the best-fit template models allowed us to estimate,
for each HDF--N galaxy, its color excess, its apparent and intrinsic UV
luminosity at 1500 \AA, and the proper cosmological distance,
according to its photometric redshift. Spectroscopic redshift was
forced in the fit whenever available.

A check on the spectroscopic sample of the HDF--N (Cohen \etal2000)
confirms that our SED fitting procedure provides a very consistent
estimate of $z$ within a relative accuracy $\Delta z/(1+z) \sim 0.04$
(see Massarotti \etal 2001b for a discussion).

\section{The comoving UV luminosity density}

The comoving luminosity density, $\rho_{(1500)}$, was computed in the
range $1 \leq z \leq 4.5$ by summing up the 1500 \AA\ contribution
from each HDF--N galaxy and adopting the $V_\mathrm{max}$ formalism
according to Lilly \etal (1996).  A flat Universe with $H_o =
50$~Km~s$^{-1}$~Mpc$^{-1}$, $q_o=0.5$ and $\Lambda=0.0$ was assumed,
throughout, in our calculations. Note that for this redshift range the
1500 \AA\ rest--frame emission is sampled by the deep HST photometry,
while the 4000 \AA\ Balmer break is still within the infrared
photometric bands of the IRIM observations.

The resulting apparent UV luminosity density at 1500 \AA\ is shown in
the second column of Table~1. A plot of its evolution with redshift is
displayed in the upper panel of Fig.~1. The nominal error bar on
$\rho_{(1500)}$ in Table~1 has been estimated by means of a
Monte-Carlo simulation, taking into account photometric errors of each
HDF--N galaxy as reported in the original catalog. This is certainly a
lower limit to our uncertainty because we are not considering
statistical scatter in the galaxy number counts and the sample
incompleteness. In the following analysis we tried however to account
for all these effects following the discussion in the current
literature. Results obtained with L99+BUZ and L99+BC models are in
agreement within the uncertainties due to photometric errors.

In Fig.~1 our results are compared with the 1500 \AA\ data of MPD98 for
$z > 2$, and with those of C97, obtained at 2800 \AA\ for $z < 2$. The
C97 data suggest a slightly higher luminosity density at lower
redshifts. One should consider, however, that dust reddening is milder
at 2800 \AA\ than at 1500 \AA.  By repeating our calculations for the
HDF--N galaxies at 2800 \AA, the resulting apparent UV density raises to
$\log \rho_{(2800)} = 26.48$ (in $\log$ units of
ergs~s$^{-1}$~Hz$^{-1}$Mpc$^{-3}$) at $z=1.5$, thus better matching
C97 estimates, the observed discrepancy being due to the different
wavelength range used.

Our values of $\rho_{(1500)}$ are larger than those of MPD98 because
of: (1) the higher efficiency of the photometric-redshift technique
compared to the $B$ drop-out galaxy selection (see Fontana \etal 2000)
and (2) the deeper magnitude limit of the sample that we used.

It is important to stress that all data in our analysis are {\it not}
corrected for the effect of the surface brightness dimming
(Pascarelle, Lanzetta, \& Fern\'andez--Soto 1998) or for sample
incompleteness in the galaxy number counts. As is widely recognized,
the latter correction is still very poorly known for high-redshift
data, and probably exceeds a factor of $1.5\div 2$ for faint galaxy
counts beyond $z > 2$ (Buzzoni 2001; MPD98).  As our main concern is
the effect of dust absorption at higher redshifts, a detailed analysis
of incompleteness in the HDF--N data is beyond the scope of this letter.

\section{The extinction--corrected luminosity density at 1500 \AA}

The UV output changes significantly when HDF--N photometry is corrected
for the galaxy color excess obtained from the SED fit. The
extinction--corrected estimates of $\rho_{(1500)}$ in different
redshift bins are shown in the third column of Table~1 and in the
lower panel of Fig.~1.  Also plotted are values usually quoted in the
literature, obtained by applying the Steidel \etal (1999) correction
to the C97 and MPD98 values.

The extinction--corrected values of $\rho_{(1500)}$ are noticeably
larger, by a factor of about $8\div 14$, than the apparent luminosity
density we obtained by directly summing up galaxy fluxes. Quite
interestingly {\it the range of our 1500 \AA\ correction factor is
much higher than previously assumed in the literature}, as summarized
in Table~2.

The reason for the striking difference from Steidel \etal (1999) does
not reside in any substantial change of the input physics (Steidel
\etal 1999 also used the Calzetti attenuation law to account for
dust absorption) but rather in a different correction procedure, that
relies on the {\it individual} value of $E(B-V)$ for each galaxy
instead of the {\it mean} value over the whole sample.

To better understand this effect consider, for instance, a toy sample of
$N$ galaxies, all with the same absolute luminosity but
different amounts of reddening, distributed around a mean value
$E(B-V)_{ave} = \sum_j E(B-V)_j/N$. The mean absorption coefficient will be 
$A_{ave} = k(\lambda)E(B-V)_{ave}$, where $k(\lambda)$ is the adopted
attenuation law. Following Steidel \etal (1999), the correction factor 
would be:
\begin{equation}
F_{ave} =  10^{0.4 k \sum_j E(B-V)_j/N} = 10^{0.4 k E(B-V)_{ave}}\,,
\end{equation}
equal for all galaxies. 
On the other hand, in our approach, the {\it individual} value of
$E(B-V)$ is applied for each galaxy, and an effective correction
factor can be defined for the whole sample as
\begin{equation}
F_{eff} = \sum_j 10^{0.4 k E(B-V)_j}/N = 10^{0.4 k E(B-V)_{eff}}\,.
\end{equation}
It can easily be demonstrated that, in general, $F_{eff} \geq
F_{ave}$ or, equivalently, that the {\it effective} color excess
$E(B-V)_{eff}$ systematically exceeds the value of $E(B-V)_{ave}$.

Estimating the energy absorbed by dust using $E(B-V)_{ave}$ over the
whole sample instead of $E(B-V)_{eff}$ can therefore result in a
significant error. 

This point can also be verified in Fig.~2, where we computed both the
mean and the effective color excess for the full HDF--N sample,
according to the different sets of galaxy reference models. Using the
L99+BUZ template set we obtain $E(B-V)_{ave} = 0.132$, while
$E(B-V)_{eff} = 0.215$.

Moreover, our analysis show that $E(B-V)$ correlates with absolute
dust--corrected UV luminosity, up to the fainter magnitudes levels
observed (see also MCH99, Adelberger \& Steidel 2000 for a similar
result on brighter spectroscopic samples).  If high intrinsic UV
luminosity starbursts have higher dust content, the underestimate
caused by using $ E(B-V)_{ave} $ is even larger, explaining the
discrepancy with Steidel \etal (1999, see also Papovich et al. 2001).

The previous considerations suggest that, while typical reddening in
high-redshift objects could be low, the total amount of energy
absorbed by dust can be considerable. This is because the total budget
of absorbed energy is dominated by the amount of UV luminosity
suppressed in high- and moderately-reddened galaxies, and these
are, in turn, the brightest UV galaxies of the sample. 

When comparing our results with the MCH99 estimates of Table~2, one
should keep in mind that their $U$ drop-out selection technique is
biased against heavily reddened galaxies. Choosing only those objects
bluer than $(V_{606}-I_{814})_\mathrm{AB}=0.5$, as in MCH99, will
cause one to miss all the galaxies detected in $I_{814}$ but with a
color excess $E(B-V) \geq 0.32 \div 0.23$ (at $z =2 \div 3.5$
respectively), and with an UV intrinsic spectral slope (\ie in the
absence of dust absorption, MCH99) $\beta_0 > -2.23$. Moreover,
and this is a crucial factor, since MCH99 assume for all starburst
galaxies $\beta_0 \sim -2.23$, the energy absorbed by dust is
underestimated for objects that fit their selection criteria but
possess UV intrinsic spectral slope $\beta_0 < -2.23$. Note that in
the redshift range $z=2.0-3.5$ we find for our chosen templates a
median value $\beta_0=-2.6$.

A safe lower limit to the amount of dust absorption in the HDF--N
galaxies can be computed by replacing the Calzetti attenuation law
with the Small Magellanic Cloud (SMC) extinction curve (Pr\'evot \etal
1984, and Bouchet \etal 1985). This curve derives from observations of
individual stars, and therefore does not include the contribution of
photon scatter along the line of sight. Consequently a lower amount of
dust is required to induce the same UV luminosity dimming.  Repeating
our calculations with the SMC extinction curve, for galaxies at
$<z>=2.75$ we obtain a value of $F_{(1500)} = 2.9$.

\section{UV vs.\ FAR--IR inferred SFRD}

A straightforward relationship exists between UV luminosity and SFR:
UV-enhanced stars of high mass can be linked to the actual star
formation in a galaxy once the initial mass function (IMF) for the
stellar distribution is assumed. As far as the comoving luminosity
density is concerned, we could therefore write

\begin{equation}
\rho_{UV} = \alpha_{({\rm IMF})}\ {{\rm SFRD}\over{{\rm M}_\odot\ {\rm
yr}^{-1} }}\ {\rm ergs\ s}^{-1}{\rm Hz}^{-1}\,,
\end{equation}

with $\alpha$ depending on the wavelength and the IMF details. Based
on the BC models, at 1500 \AA\ and for a Salpeter IMF, MPD98 suggest
$\alpha_{1500} = 8.0\ 10^{27}$ ergs~s$^{-1}$~Hz$^{-1}$.  This is in 
good agreement with the corresponding calibration that depends on the 
Buzzoni (2001) models which provide $\alpha_{1500} = 8.6\ 10^{27}$
ergs~s$^{-1}$~Hz$^{-1}$.

Optically--selected galaxies in the redshift range between $1 \lesssim
z \lesssim 4.5$ probably give a marginal contribution to the bright
source counts at far--IR/submillimeter wavelengths. Only two HDF--N
galaxies from the Fern\'andez-Soto, Lanzetta, \& Yahil (1999) catalog,
within this redshift range, might be tentatively associated to 850
$\mu m$ SCUBA sources at a flux threshold of $S_{850}> 2$ mJy (Hughes
\etal 1998).

It is however relevant to compare the galaxy star formation history
inferred from the observations in the two spectral windows, in order
to estimate the energy contribution of optically--selected high
redshift starburst galaxies to the sub--millimeter background at faint
flux levels $S_{850}< 2$ mJy, corresponding to $\sim 75 \%$ of the
total energy budget (BCR00). This will give us further clues to
assess, independently, the self-consistency of our galaxy evolutionary
scenario.

The cosmic SFRD, as derived from the (extinction--corrected)
$\rho_{(1500)}$ estimates of Table~1, is displayed in Fig.~3. Solid
dots show our results while stars are BCR00 results from SCUBA
observations.  The BCR00 data have been transformed to our adopted
cosmological model, and completeness corrections have been applied
according to Fixsen \etal (1998).

We stress again that, owing to the HDF--N sample incompleteness, our
results should certainly be taken as safe {\it lower} limits. Quite
comfortingly, however, our new approach to dust correction of the UV
data substantially improves the match with the far--IR observations,
suggesting that the bulk of far--IR energy budget at faint flux levels
of $S_{850} < 2$ mJy is produced by galaxies detected in optical HST
data.

\acknowledgments It is a pleasure to thank Roberto Della Ceca and
Paolo Saracco for useful discussions and Jack Sulentic for carefully
reading the manuscript and giving his suggestions. This project
received partial financial support from Fondazione Cariplo and from
COFIN 00-02-016 grant.

\clearpage

\begin{deluxetable}{ccc}
\tabletypesize{\small}
\tablecolumns{3}
\tablewidth{0pc}
\tablecaption{Comoving UV luminosity density}
\tablehead{
\colhead{Redshift} & \multicolumn{2}{c}{$\rho_{1500}$\tablenotemark{(a)}}\\
\colhead{z} & \colhead{apparent} & \colhead{extinction} \\
\colhead{}  & \colhead{}         & \colhead{corrected}
}
\startdata
1.0-2.0 & $2.1\pm0.1$ & $17\pm 1$ \\
2.0-3.5 & $2.3\pm0.1$ & $28\pm 3$ \\
3.5-4.5 & $1.0\pm0.1$ & $12\pm 2$ \\
\enddata
\tablenotetext{(a)}{Flux density at 1500 \AA\ is given in
unit of $10^{26}$ ergs~s$^{-1}$~Hz$^{-1}$Mpc$^{-3}$.}
\end{deluxetable}

\clearpage

\begin{deluxetable}{cll}
\tabletypesize{\small}
\tablecolumns{3}
\tablewidth{0pc}
\tablecaption{UV correcting factor for dust absorption}
\tablehead{
\colhead{Mean redshift} & \colhead{$F(1500)$\tablenotemark{(a)}} & \colhead{Reference} \\
}
\startdata
3.04 & 4.7           & Steidel et al \\
2.75 & 5.4 $\pm$ 0.9 & MHC99 \\
2.75 & 12 $\pm$ 2    & This work      \\
\enddata
\tablenotetext{(a)}{$\rho_{(1500)}$ (extinction--corrected) = $\rho_{(1500)}$ (apparent) $\times F(1500)$}
\end{deluxetable}

\clearpage

\begin{figure}
\plotone{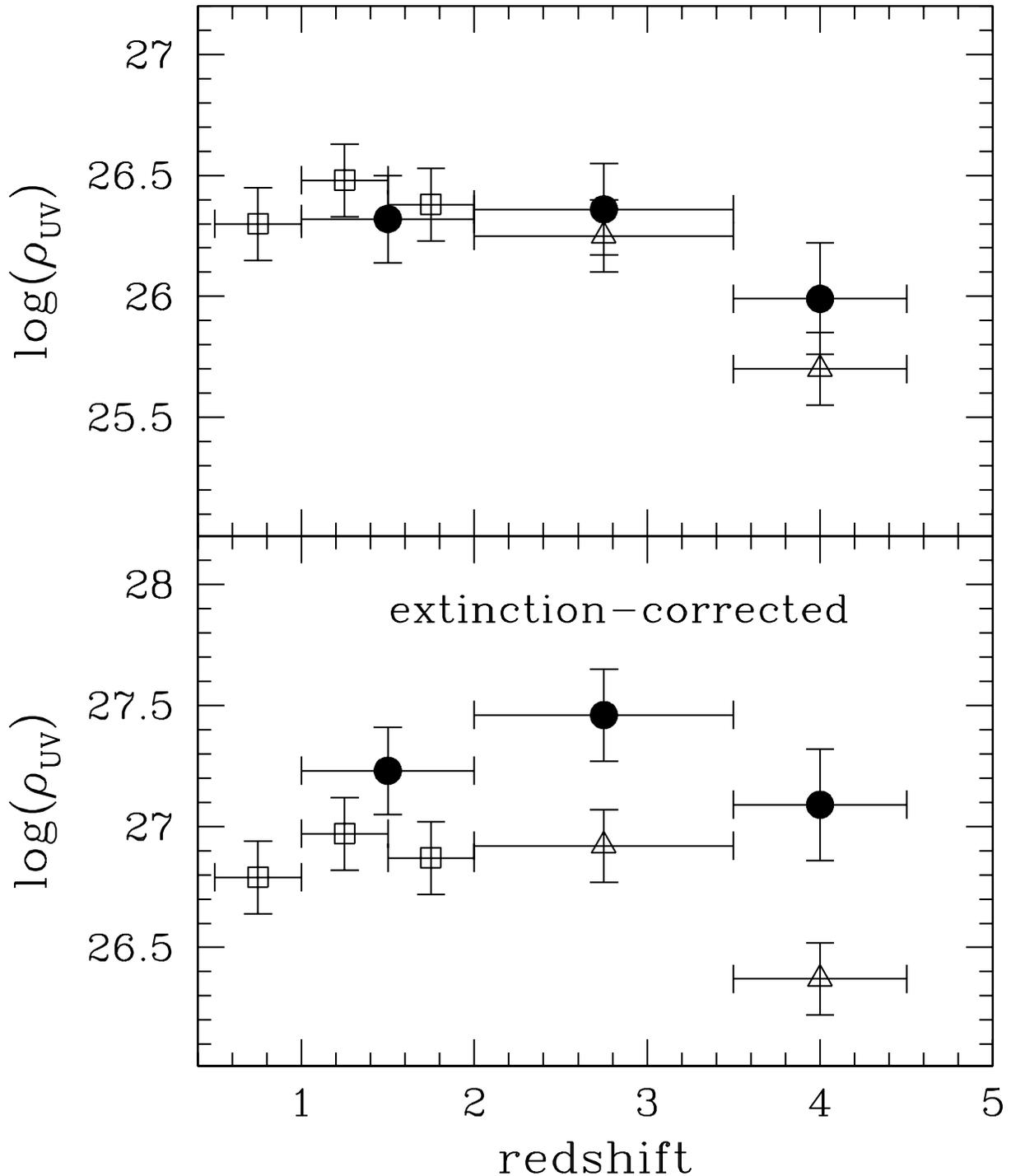}
\caption{Redshift evolution of the comoving UV luminosity density (in
$\log$ unit of ergs~s$^{-1}$~Hz$^{-1}$Mpc$^{-3}$), as sampled by the
HDF--N galaxies.  The apparent value of $\rho_{(1500)}$ at 1500 \AA,
from the data in Table~1, is displayed in the upper panel, while its
trend after correction for dust absorption is reported in the lower
panel. Our results (solid dots) are compared with the MPD98 1500\AA\
data at $z >2$ (triangles), and with the 2800 \AA\ estimates of C97
at $z < 2$ (squares).  Error bars on the points try to account for a
more realistic error estimates including field--to--field variations
(see Fontana \etal1999) and the errors induced by the unknown
incompleteness of the sample studied.}
\end{figure}

\clearpage

\begin{figure}
\plotone{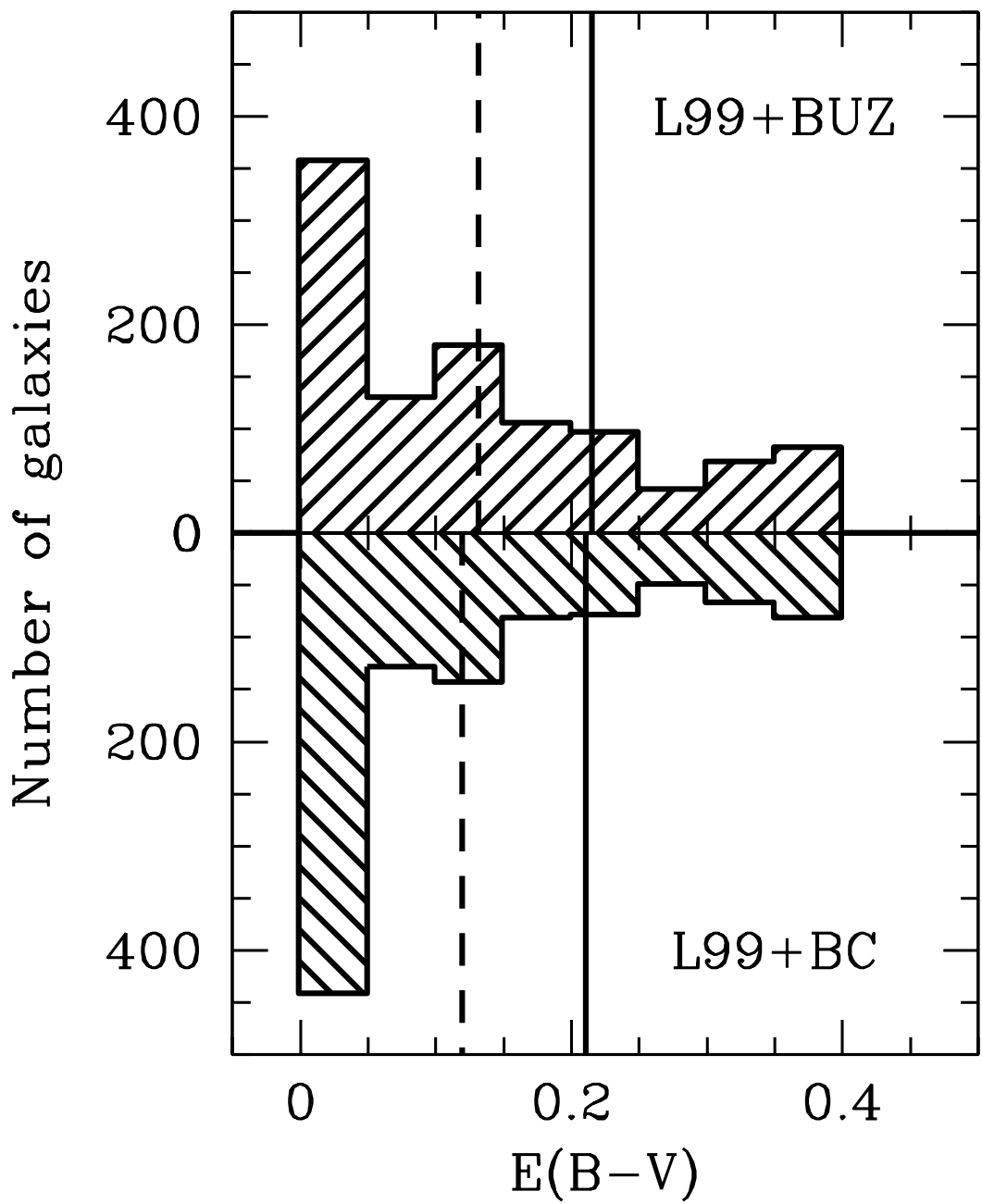}
\caption{Color excess distribution of the HDF--N galaxies, according to
our SED fitting comparing the output of the L99+BUZ and L99+BC
reference sets of galaxy templates.  The best-fit value of $E(B-V)$ for
the 1067 galaxies in the sample is displayed together with the
computed values of $E(B-V)_{ave}$ (dashed lines) and
$E(B-V)_{eff}$ (solid lines), as discussed in the text.}
\end{figure}

\clearpage

\begin{figure}
\plotone{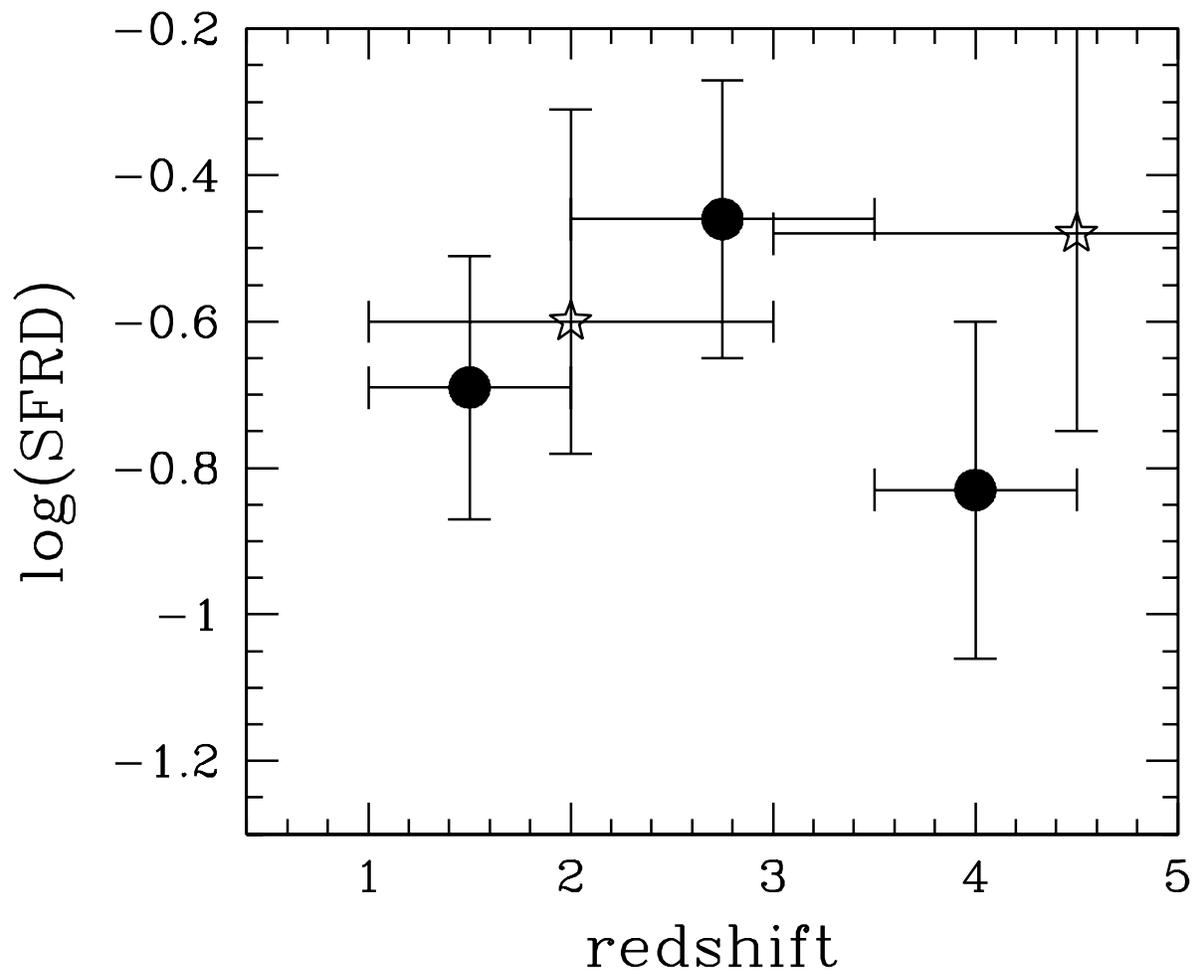}
\caption{Inferred cosmic SFRD according to the evolution of the
extinction--corrected comoving UV luminosity density, $\rho_{(1500)}$,
from the data of Table~1 (solid dots). Our calibration is from
eq.~(3), with $\alpha_{1500} = 8.3\ 10^{27}$ ergs~s$^{-1}$~Hz$^{-1}$.
Once accounting for dust absorption, our data from the HDF--N galaxies
are consistent with the SCUBA far--IR background estimates from BCR00
(stars) indicating a non-decreasing SFRD at least up to $z \sim 3$.}
\end{figure}

\end{document}